\documentclass[11pt]{article}
\usepackage{amsmath}
\usepackage{graphicx}
\usepackage[usenames,dvipsnames,table]{xcolor}
\usepackage{amssymb}
\usepackage{amsmath}
\usepackage{float}
\usepackage{scrtime}
\usepackage{fancyhdr}
\usepackage{subfigure}
\usepackage{authblk}
\usepackage{lipsum}
\usepackage{rotating}
\usepackage{floatpag}
\usepackage{varioref}
\usepackage{slashed}
\usepackage{enumerate}
\usepackage{listings}
\usepackage{cancel}
\usepackage{wrapfig}
\usepackage[margin=1.1in]{geometry}

\usepackage{caption}
\DeclareCaptionFont{white}{\color{white}}
\DeclareCaptionFormat{listing}{%
  \parbox{\textwidth}{\colorbox{gray}{\parbox{\textwidth}{#1#2#3}}\vskip-4pt}}
\usepackage[colorlinks=true
,urlcolor=blue
,anchorcolor=blue
,citecolor=blue
,filecolor=blue
,linkcolor=black
,menucolor=blue
,pagecolor=blue
,linktocpage=true
]{hyperref}

\usepackage[numbers,sort&compress]{natbib}

\setlipsumdefault{131}

\newcommand{\code}[1]{{\small{\texttt{#1}}}}

\labelformat{section}{Section #1}
\labelformat{subsection}{Section #1}
\labelformat{equation}{Eq.~(#1)}
\labelformat{figure}{Fig.~#1}
\labelformat{subfigure}{Fig.~\thefigure#1}
\labelformat{table}{Tab.~#1}
\labelformat{lstlisting}{Listing~#1}

\def\tchi{\ensuremath{\widetilde{\chi}}}
\def\mcha{\ensuremath{m_{\widetilde{\chi}_1^+}}}
\def\mneu{\ensuremath{m_{\widetilde{\chi}_1^0}}}

\def\be{\begin{equation}}
\def\ee{\end{equation}}
\def\bea{\begin{eqnarray}}
\def\eea{\end{eqnarray}}

\newcommand{\met}{E_{\rm T}^{\rm miss}}

\graphicspath{{./figures/}}

\begin{document}

\floatpagestyle{plain}

\pagenumbering{roman}

\renewcommand{\headrulewidth}{0pt}
\rhead{
OHSTPY-HEP-T-14-004\\
}
\fancyfoot{}

\title{\huge \bf{Degenerate gaugino mass region and mono-boson collider
signatures}}

\author[$\dag$]{Archana Anandakrishnan}
\author[$\dag$]{Linda M. Carpenter}
\author[$\dag$]{Stuart Raby}

\affil[$\dag$]{\em Department of Physics, The Ohio State University,\newline
191 W.~Woodruff Ave, Columbus, OH 43210, USA \enspace\enspace\enspace\enspace
\medskip}

\maketitle
\thispagestyle{fancy}

\begin{abstract}\normalsize\parindent 0pt\parskip 0pt
In this paper we discuss search strategies at the LHC for light electroweak gauginos which are mostly Wino-like, Higgsino-like or an admixture. These states are typically degenerate with decay products that are less energetic and hence difficult to detect. In addition, their production cross-sections at a hadron collider are suppressed compared to colored states such as the gluinos. In order to detect these states one needs to trigger on initial or final state radiation. Many previous analyses have focussed on mono-jet and mono-photon triggers. In the paper we argue and show that these triggers are unlikely to succeed, due to the large background from QCD backgrounds for the mono-jet searches and the fact that the $p_T$ distribution of the mono-photons are rapidly decreasing functions of $p_T$.  We show this with both an analytic calculation of photons in the initial state radiation and also a detailed numerical analysis. We then argue that mono-Z triggers, from Z decaying into charged leptons may well provide the best search strategy, in particular for Higgsino-like and mixed cases.
\end{abstract}

\clearpage
\newpage

\pagenumbering{arabic}

\section{Introduction}

Electroweak gauginos and higgsinos, in most of the supersymmetric model-space are the most likely candidates for the spectrum's Lightest Supersymmetric Particle (LSP) or Next-to Lightest Supersymmetric Particle (NLSP).  Moreover in many appealing models, the lightest chargino and neutralino, either Wino of Higgsino-like are quite mass degenerate. Examples of models exhibiting degenerate spectra include minimal versions of anomaly mediation~\cite{Giudice:1998xp, Randall:1998uk}, mirage mediation~\cite{Choi:2007ka, Lowen:2008fm, Anandakrishnan:2013cwa}, light higgsino natural supersymmetry models~\cite{Baer:2012up} and Higgsino-world scenarios~\cite{Kane:1998ib}. As the first kinematically accessible states, it is important to create experimental searches which will be sensitive to these particles. However in the mass degenerate scenario searches for electroweakinos (ewkinos) become quite hard, involving non-standard topologies as displaced vertices, kinks and disappearing tracks.

While supersymmetry (SUSY) remains the leading candidate for physics beyond the Standard Model, current searches at the LHC have not yet revealed supersymmetric particles. Minimal versions of leading supersymmetry breaking communication schemes such as mSUGRA, gauge mediation and anomaly mediation predict relatively similar particle spectra with the heaviest sparticles being squarks, roughly an order of magnitude heavier than the lightest supersymmetric particles, the ewkinos. In such models, the hope was that the smoking gun signal for supersymmetry would be in jets $+$ missing energy channel from the strong production of pairs of gluinos or squarks. In view of models with maximal naturalness, these particles were hoped to be under 1 TeV in mass. Current constraints, however, are pushing us to look for SUSY in different places. Current bounds from ATLAS and CMS have pushed mass bounds for squarks and gluinos, which decay into typical jets $+$ missing energy channels, into the 1-2 TeV range~\cite{Aad:2014wea}. Separately, the Higgs mass constraint of 126 GeV~\cite{Aad:2012tfa,Chatrchyan:2012ufa} hints that squark masses should reside in the multi-TeV range in order to facilitate large loop contributions to the tree level Higgs mass. Reasonable model space exists where squarks may be in the 10 TeV range. Yet the LHC 5-sigma discovery potential for gluinos is less than 2 TeV~\cite{Ulmer:2013csa}.  In light of the possibility of spectra with heavy colored sparticles, one must reconsider the channels in which supersymmetry is most likely to make its first appearance, and the ewkinos become an important discovery channel for SUSY.

Topologies employing mono-boson final states have been quite useful for excluding parameter regions of  many Beyond the Standard Model (BSM) scenarios. In particular mono-jet searches have been a useful tool in constraining various scenarios from dark matter models to extra-dimensional models~\cite{Goodman:2010ku, CMS-PAS-EXO-12-048, ATLAS-CONF-2012-147}. The success of the mono-jet searches have led to analysis expansions into mono-photon, mono-W/Z, and mono-Higgs final states to constrain new physics beyond the standard model~\cite{Aad:2012fw, Carpenter:2012rg, Carpenter:2013xra, Aad:2014vka}. See also~\cite{Chen:1999yf, Baer:2009dn, Giudice:2010wb, Fox:2011fx, Rajaraman:2011wf, Fox:2011pm, Han:2013usa, Han:2013kza, Baer:2014cua, Berlin:2014cfa}.

In this paper we examine the effectiveness of mono-boson search channels for the discovery of mass degenerate ewkinos. We show that the massive gauge boson channels, in particular the mono-Z final state have a 5-sigma discovery potential for ewkinos in the 13 TeV run at the LHC. We find that massless gauge-boson final state searches such as the mono-photon fail to reach sensitivity to this mass degenerate SUSY scenario, and we give a quantitative analytic explanation of this failure using simple effective operator techniques. The large QCD backgrounds to the mono-jet searches also make this a less efficient search strategy.

This paper continues as follows, in~\ref{sec:spectrum}, we discuss the parameters and mass spectrum of degenerate ewkinos and in~\ref{sec:monob} we present the sensitivities of mono-photon and mono-jet analyses to our SUSY scenario. In~\ref{sec:monoz} we discuss the discovery potential of the mono-Z search channel and briefly discuss non-trivial topologies which would be quite powerful in the degenerate ewkino mass region in~\ref{sec:other}. Finally in~\ref{sec:conclusion}, we conclude.

\section{The electroweakino spectrum}
\label{sec:spectrum}
The masses of the electroweakino sector are fixed by the Majorana mass parameters of the pure bino and wino $M_1$ and $M_2$, and also by the $\mu$ term, and tan$\beta$. In minimal scenarios such as mSUGRA, or minimal gauge mediation, and simple versions of anomaly mediation $M_1$ and $M_2$ do not vary independently. More generally however the ratio between $M_1$ and $M_2$ can vary~\cite{Choi:2007ka, Lowen:2008fm, Carpenter:2008he, Anandakrishnan:2013cwa}. A wide region of MSSM parameter space exists with mass degenerate chargino and neutralino,  all that is needed is that the lightest chargino is Wino or Higgsino like.  Here we consider three benchmark scenarios:

\begin{itemize}
 \item $M_2 < M_1 < \mu $ In this limit, the LSP and the NLSP are pure wino.
 \item $M_2 \simeq \mu << M_1$ In this limit, the LSP and the NLSP are a wino-Higgsino mixture.
 \item $\mu \simeq M_2 \simeq M_1$ In this limit, the LSP and the NLSP are  pure Higgsino.
\end{itemize}

The mass difference between the LSP and the NLSP at tree level is set by the three parameters above and $\tan \beta$~\cite{Cheng:1998hc}:
\be
\Delta M_{\rm tree} = m_{\tilde{\chi}_1^+} - m_{\tilde{\chi}_1^0} = \frac{M_W^2}{\mu^2}\frac{M_W^2}{M_1 - M_2} \tan^2 \theta_W \sin^2 2\beta +
          \mathcal{O} \left( \frac{1}{\mu^3} \right) . \label{deltam}
\ee

\begin{table}[h!]
\begin{center}
\renewcommand{\arraystretch}{1.0}
\scalebox{0.93}{
\begin{tabular}{|c|c|c|c|c|c|c|c|}
\hline
Point & $\mu$ & $M_2$ & $M_1$  & \mcha & \mneu & $\Delta M$ & $\tau_{\tilde{\chi}^\pm}$\\
\hline
Wino 1(98\% Wino) & 700 & 100 & 200 & 98.07 & 98.06  & 0.152 & 92.8216 \\
Higgsino 1(98\% Higgsino) & 600 & 300 & 3000 & 292.27 & 292.26 & 0.178 & 39.1402 \\
Wino 2 (96\% Wino) & 540 & 150 & 180 & 145.71 & 145.54  & 0.321 & 4.3766 \\
Higgsino 2 (88\% Higgsino) & 150 & 300 & 1200 & 136.38 & 130.26 & 6.29 & $1.98665 \times 10^{-5}$\\
Mixed 1( 72\% Wino and 23 \% Bino) & 500 & 200 & 200 & 193.22 & 191.25 & 2.12 &0.00241474 \\
Mixed 2( 65\% Wino and 23 \% Bino) & 360 & 200 & 200 & 186.13 & 182.21 & 4.07 & $7.06516 \times 10^{-5}$\\
Mixed 3( 28\% Wino and 23 \% Bino) & 180 & 200 & 200 & 138.27 & 127.75 & 10.68 & $5.30822 \times 10^{-7}$\\
\hline
\end{tabular}
}
\caption{\label{tab:Benchmarks} We compile a series of benchmark points for Wino-like, Higgsino-like and well-mixed chargino scenarios to demonstrate mass splittings and chargino lifetimes.  $\tan \beta = 30$ is fixed.}
\end{center}
\end{table}

In addition, there are 1-loop electroweak corrections to the chargino masses that make the charginos heavier by about 150 MeV. The mass splitting of the lightest chargino and neutralino ($\Delta M = \Delta M_{\rm tree} + \Delta M_{\rm 1-loop}$) is thus always greater than the pion mass. We will consider benchmark points covering all three scenarios and a compilation of benchmark points can be found in~\ref{tab:Benchmarks}. It is clear from~\ref{tab:Benchmarks} that as $\mu$ increases, the mass difference between the LSP and the NLSP decreases. Therefore, the wino-like benchmarks predominantly have very small mass splittings. For the purely higgsino-like scenario, the mass dependence closely depends on the difference between $M_1$ and $M_2$. The tree-level mass difference is $\tan \beta$ suppressed. So at large $\tan \beta$, the splitting is smaller. We have fixed $\tan \beta = 30$ in our benchmark points. The production cross-sections for the different regimes has been discussed in detail in Ref.~\cite{Han:2013kza}.

For small mass splitting, the decay of the lightest chargino proceeds through an off-shell W, $\tchi^{\pm}\rightarrow W^{*}\tchi^0$, dominated by hadrons down to $\Delta M < 10$ GeV and by pions below $\Delta M < 1$ GeV.  Further, for mass splitting near the pion mass, the chargino decay width is sufficiently small to produce cm sized displaced vertices~\cite{Chen:1999yf}. Smaller electroweakino mass splittings present a challenge to detectability since the decay products do not carry enough momentum to pass the triggers of the experimental searches. Standard search topologies involve isolated hard leptons from chargino decay and will fail in this parameter regime, and new event topologies must take their place.

\section{Testing Topologies : $\tchi \tchi$ + Mono Boson Final State}
\label{sec:monob}
In the pair production of ewkinos given by $pp\rightarrow \tchi^0 \tchi^0, \tchi ^{\pm}\tchi^{0}, \tchi^{+}\tchi^{-}$, the decays will be characterized as pure missing energy ($\met$), $W^{*} + \met$, and $W^{*}W^{*} + \met$ respectively. The decay products of charginos in mass degenerate scenarios are extremely soft (less energetic), and therefore both charginos and neutralinos may appear purely as missing energy in simple searches. Due to their appearance as pure missing energy, the detection of the pair produced ewkinos require a trigger on some recoiling state. One promising search topology requires triggering on a gauge-boson emitted as initial state or final state radiation. We note that for $\Delta M \lesssim 10$ GeV, the mono-boson plus + $\met$ channels will be a viable search strategy. For $\Delta M <$ 1 GeV, charginos are long lived and options such as disappearing tracks also become viable search channels. We will briefly comment on other topologies in this region of parameter space in~\ref{sec:other}. However, we will first discuss the mono-boson + $\met$ search channels which are the most promising signatures for $1 < \Delta M \lesssim 10$ GeV.

\begin{figure}[h!]
\begin{center}
\subfigure[\footnotesize Wino LSP]{
\includegraphics[width=7.68cm]{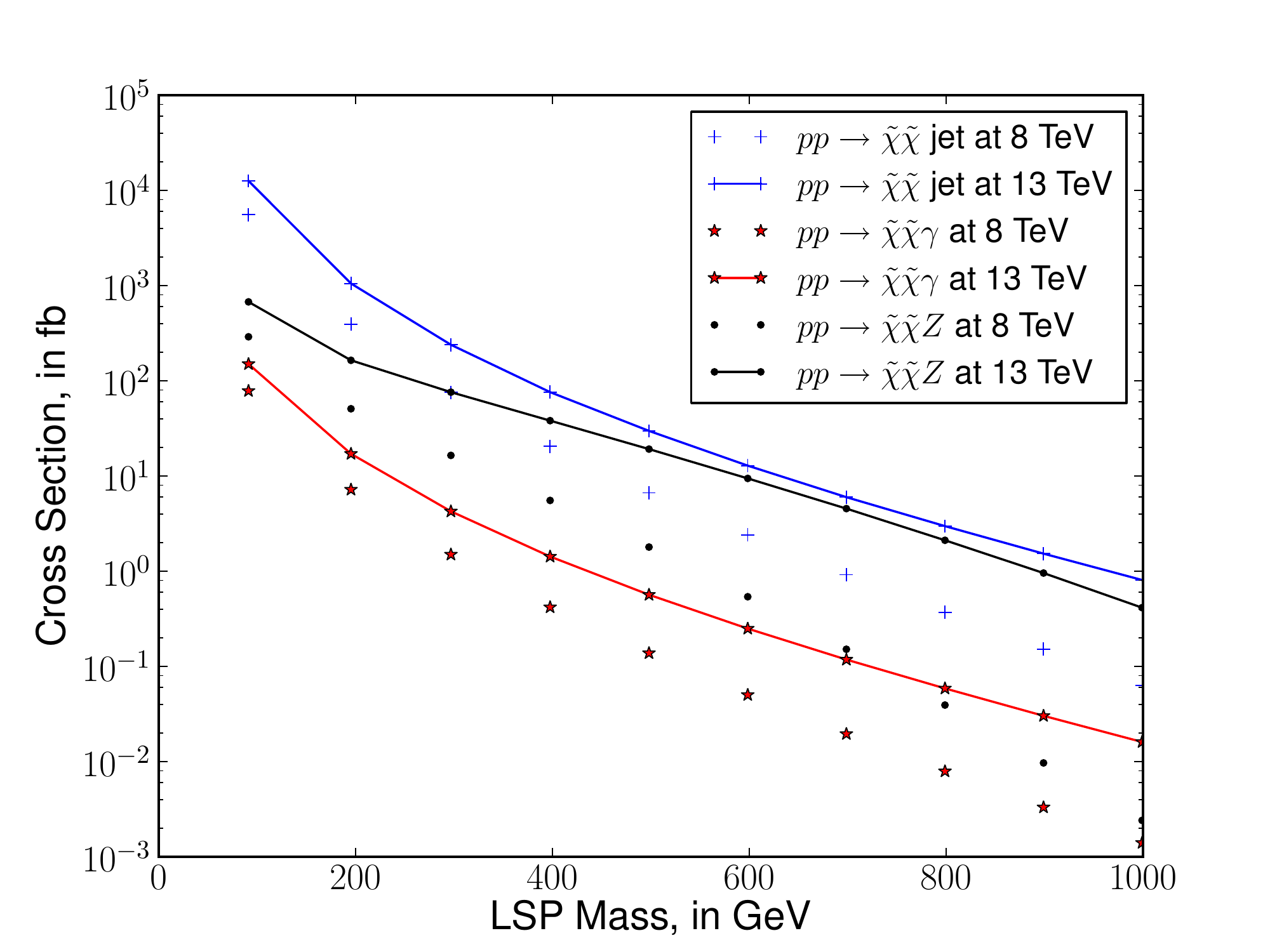}
}
\subfigure[\footnotesize Higgsino LSP ]{
\includegraphics[width=7.68cm]{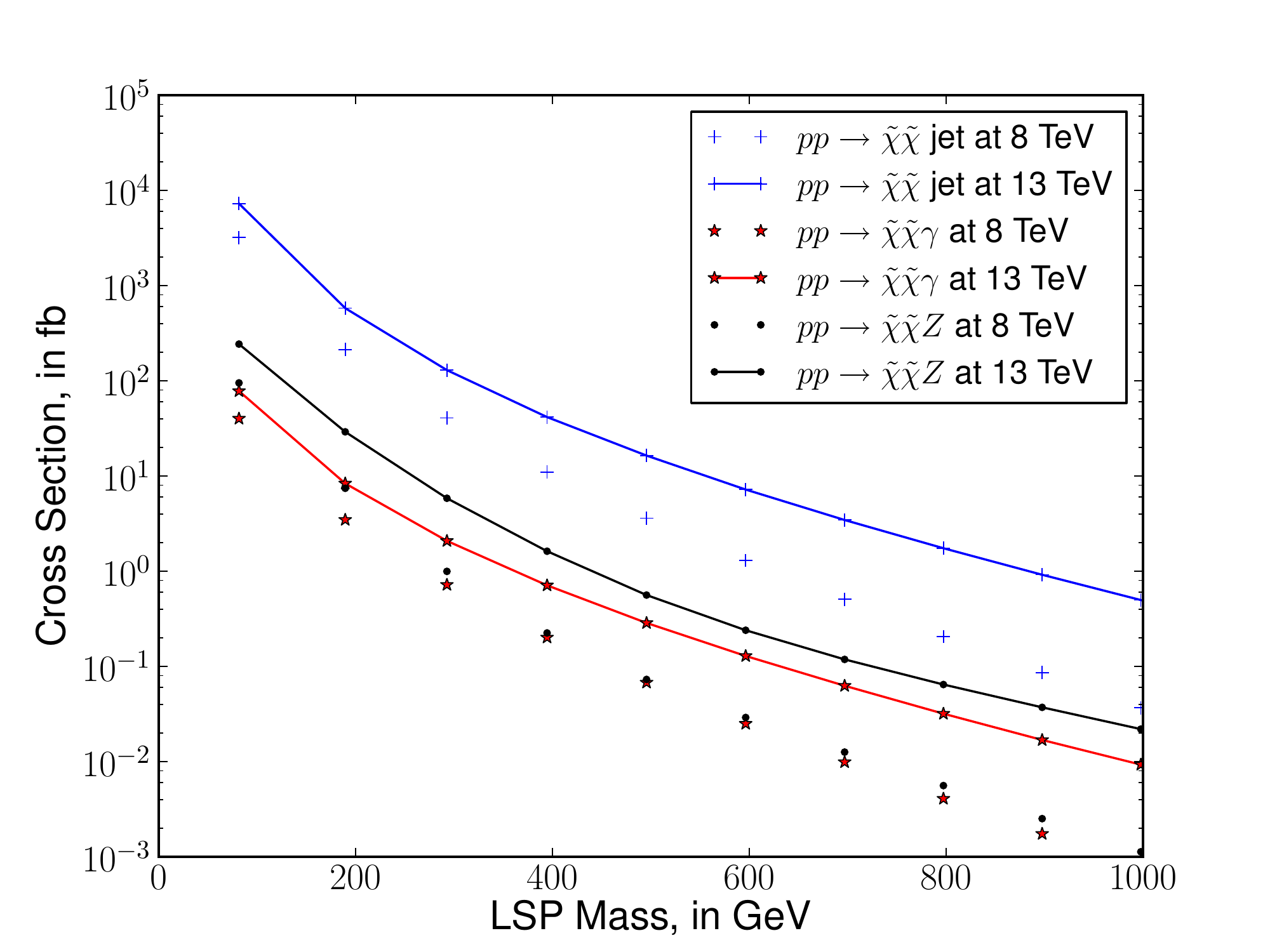}
}
\newline
\subfigure[\footnotesize Mixed wino-higgsino LSP]{
\includegraphics[width=9cm]{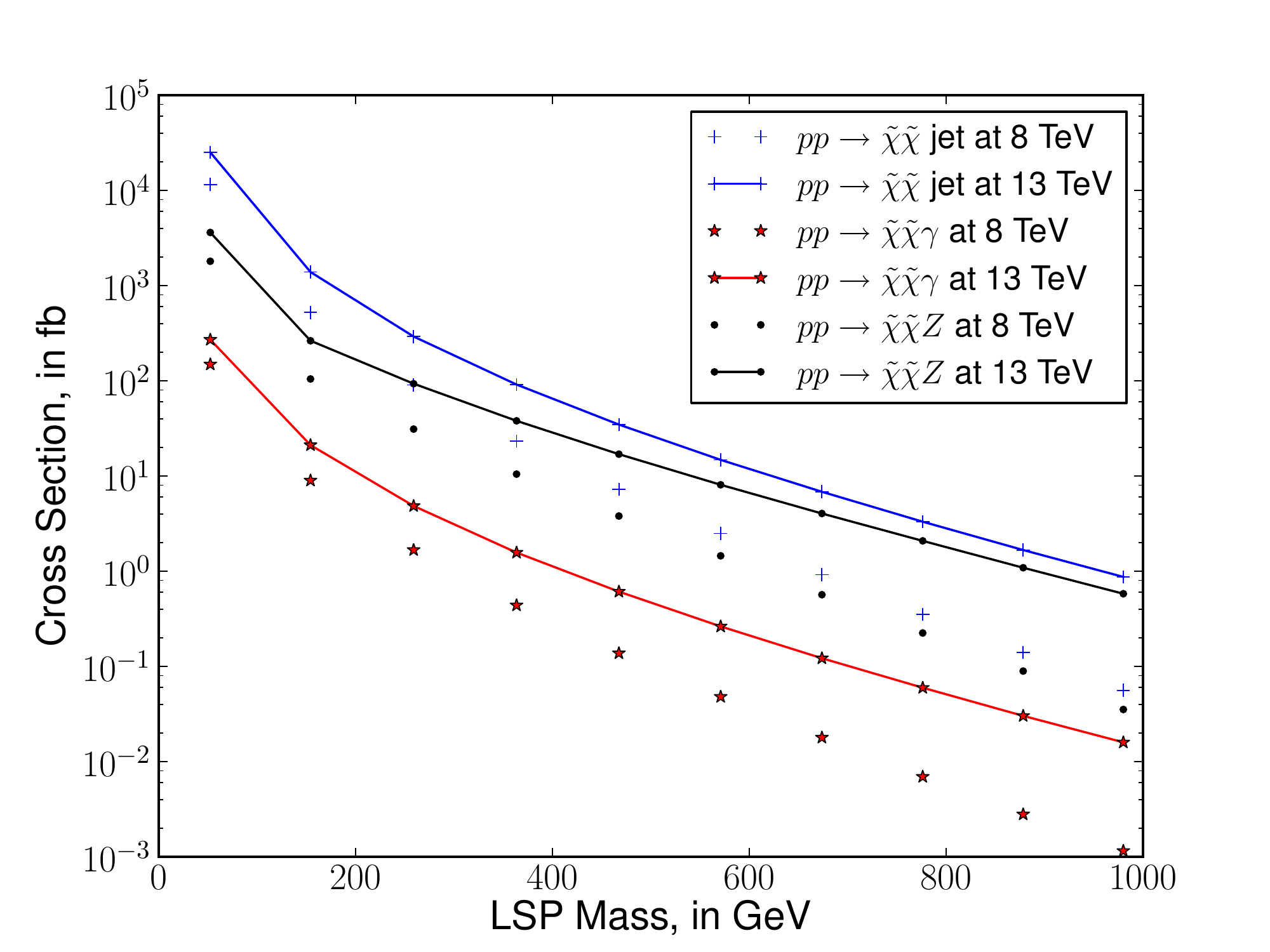}
}
\caption{Pair production cross-sections in association with a jet (${\color{blue} +}$), $\gamma$ (${\color{red} *}$), Z ($ \centerdot$) for the three scenarios discussed in the text. The production cross-sections are shown both at 8 TeV and 13 TeV center of mass energies at the LHC.}
\label{xsec}
\end{center}
\end{figure}

We consider below three possible mono-boson event topologies, $pp \rightarrow {\rm jet}+ \met$, $pp \rightarrow \gamma+ \met$, $pp \rightarrow Z+ \met$, or events with a mono-jet, mono-photon, or mono-Z.  Since both charginos and neutralinos appear to these searches as missing energy, we consider the total production cross section of a mono-boson plus all pairs of electroweakinos. Typically, the lighter states provide the largest contributions to the total cross-section. For the winos, this includes only the lightest neutralino and chargino whereas for the higgsinos, there are two neutralino states that are degenerate. We show in~\ref{xsec}, the mass dependence of the total production cross section for Wino-like, Higgsino-like, and mixed ewkino pairs for the mono-jet, mono-photon, and mono-Z final states. These production cross-sections were calculated using \code{MadGraph5} for both 8 and 13 TeV LHC center of mass energies. The pair-production cross sections are dominated by $\tchi^{+}\tchi^{-}$ and  $\tchi^{\pm}\tchi^{0}$, with $\tchi^{0}\tchi^{0}$ pairs contributing only a small amount to the total production. The production cross-sections differ significantly with ewkino content. Pure wino states have slightly higher cross-sections for associated jet, photon and Z production than the pure Higgsino-like pairs. For example, the total cross sections for the process $p p \rightarrow \tchi \tchi + Z$ at 13 TeV center of mass energy for a 100 GeV LSP is about 600 fb for pure wino states, whereas they are only about 200 fb for the pure Higgsino-like states. Mixed states exhibit the highest overall production cross sections.

We start by exploring the effectiveness of searches for discovering the degenerate ewkinos in the mono-photon, mono-jet and mono-Z final states. We attempt to recast existing CMS and ATLAS searches from the first run of LHC and in addition, we examine the sensitivity of the searches at the 13 TeV run. It is useful to define the simple criteria of sensitivity as $S/\sqrt{B}$ where S and B are the expected signal and background events. For a search to be sensitive to the existence of new physics, we require that $S/\sqrt{B}$ be at least greater than 2. The discovery potential can be defined as follows~\cite{Baer:2009dn}:
\be
S \ge \text{max} \left[ 5 \sqrt{B}, 5, 0.2 B \right] \label{sensitivity}
\ee
The three cases in~\ref{sensitivity} are for low background and small statistics, no background, and high background respectively. We assess the mono-jet, mono-photon and mono-Z search channels and estimate the search sensitivity for the three channels. Since the cross-section drops off quite rapidly with increasing mass as shown in~\ref{xsec}, we evaluate the sensitivities for the Wino 1~\footnote{We comment about the disappearing track search for this benchmark point in~\ref{sec:other}.}, Higgsino 2, and Mixed 3 benchmark points from~\ref{tab:Benchmarks}. Later, we will extend the search sensitivities to heavier masses for the 13 TeV LHC.   	

\subsection{Effective Operators: Analytic exploration of mono-photon and mono-jet events}
Recent work using event generation and cut techniques have demonstrated the failures of mono-jet and photon analyses to detect Higgsinos in the 8 TeV data set~\cite{Han:2013usa,Baer:2014cua} as well as prospects for the 13 TeV run. Below we not only broaden this statement to all cases of general ewkinos, but also present a concise kinematic argument about the limitations of searches relying on hard cuts for initial (or final) state radiation.

We consider analytically the pair production of neutralinos and charginos along with a photon in the initial state radiation (ISR). To simplify the calculation, we consider the coupling of neutralinos ($\tchi^0$) to quarks as a dimension-6 effective operator with a weak scale cut-off.

\bea
\label{eq:SMSinglet}
\mathcal{L}_{\rm eff}= \frac{1}{\Lambda^2} \tchi \tchi ~ \bar{q} q
 \nonumber
\eea

Using this effective operator we attempt to capture the important features of the physics of the production of ewkinos through an S channel electroweak gauge boson in conjunction with photon ISR. The ISR photon is emitted from either of the incoming quarks. Using the above effective operator, we calculate the analytic form of the matrix element squared:
\be
|{\emph M}|^2 =16 \frac{e^2}{\Lambda^4}(p \cdot p'-m_{\tchi}^2)\frac{1} {\sin^2 \theta_\gamma}
\ee

\begin{figure}[h!]
\begin{center}
\includegraphics[width=12cm]{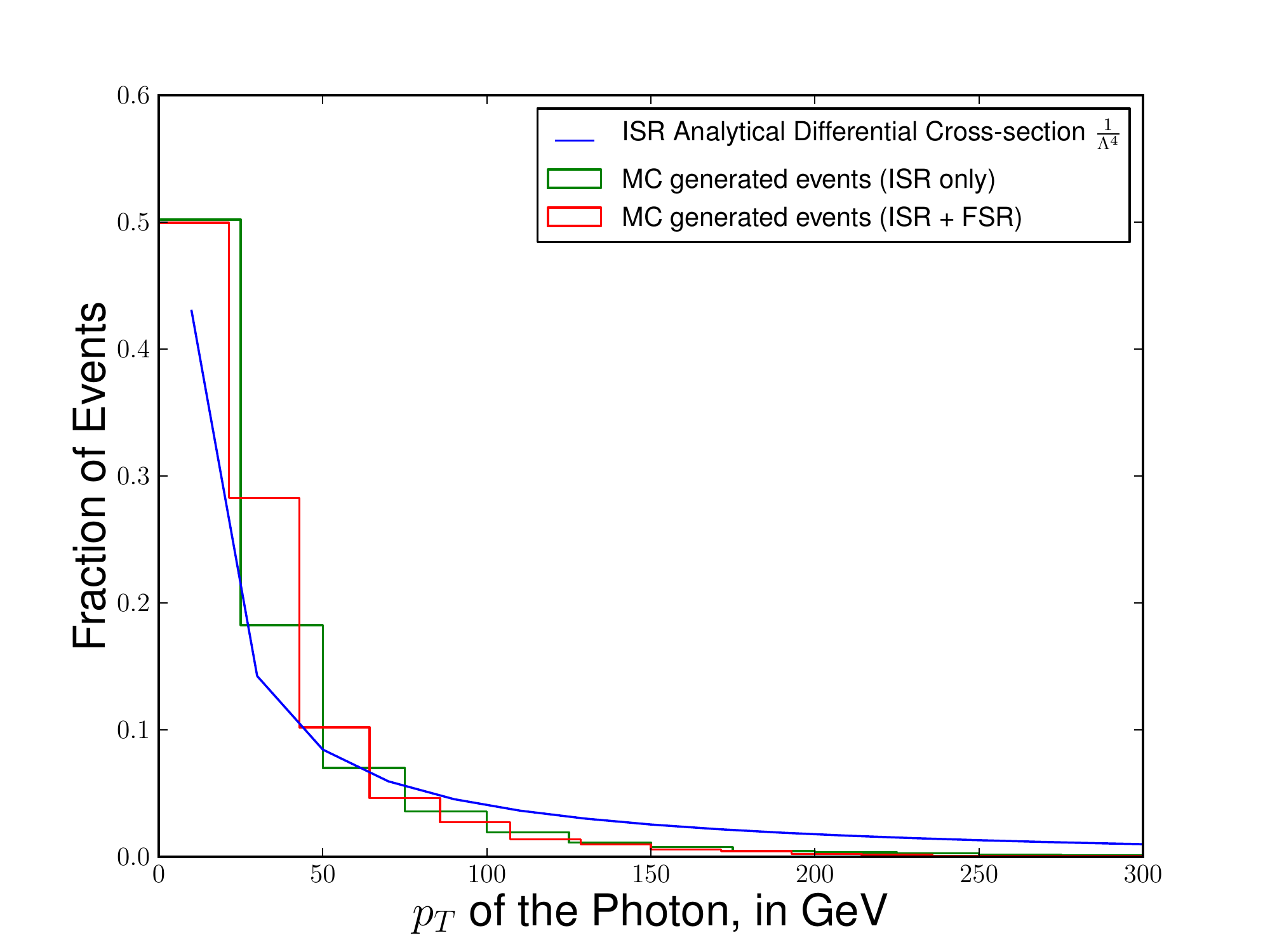}
\caption{Fraction of events binned by $p_T^\gamma$ calculated using the effective operator in~\ref{eq:SMSinglet} as well as Monte Carlo generated. The simple effective operator matches well with the $p_T$ dependence from the Monte Carlo generated events.}
\label{fig:gammapT}
\end{center}
\end{figure}

Here $p$ and $p'$ are the momenta of the \tchi s in the final state and $\theta_\gamma$ is the angle between the photon and the axis of collision. The amplitude has a velocity suppression factor due to the mass of the on-shell ewkinos. Most importantly, this matrix element squared has a collinear divergence. The cross section is maximized when the photon radiated in the initial state is in the same direction as the incoming quark, that is when the ISR photon goes down the beam pipe and the angle $\theta_\gamma = 0 \ {\rm or} \ \pi$. To determine the dependence of the total production cross section ($p p \rightarrow \tchi \tchi \gamma$) on photon $p_T^\gamma$, one must convolve the amplitude-squared with the three body phase space integral and in addition with the parton distribution functions (PDF). The photon has a minimum energy of $E_{\gamma {\rm min}} = 0$, and a maximum energy set by phase space of $E_{\gamma {\rm max}} = \frac{s-4 m_{\tchi}^2}{2 \sqrt{s}}$.  However, we are interested in the cross-section as a function of the photon $p_T^\gamma$ which is defined as:
\be
p_T^\gamma = E_\gamma \sin \theta_\gamma .
\ee
For each value of the photon momentum, the angle $\theta_\gamma$ is also bounded by phase-space between $\left[\sin^{-1} \left(\frac{p_T^\gamma}{E_{\gamma {\rm max}}} \right), \pi - \sin^{-1} \left(\frac{p_T^\gamma}{E_{\gamma {\rm max}}} \right)\right]$. We evaluate this cross-section numerically by convoluting the PDFs from~\cite{Lai:1999wy}. In~\ref{fig:gammapT}, we plot (blue line) the momentum dependence of the cross-section using the analytical expression obtained from the effective operator normalized by the total cross-section. The cross section is maximal at low values of $\theta_\gamma$ due to the collinear divergence. We see that the total cross section drops rapidly with increasing $p_T^\gamma$. The normalization of the cross-section also removes most of the dependence on the scale of the effective operator.

\ref{fig:gammapT} also shows the number of mono-photon + $\met$ events binned by $p_T^\gamma$ Monte Carlo (MC) generated using \code{MadGraph5} and showered with \code{Pythia} and \code{PGS} for the process $p p \rightarrow \gamma + \tchi^0 \tchi^0$. Both the analytical and the Monte Carlo generated results are overlaid on a single plot for a specific benchmark scenario with a $\tchi^0$ mass of 130 GeV. The Monte Carlo generated events are scaled by the total number of events such that the quantity being compared in the plot between the analytical and Monte Carlo calculation is the fraction of events in each $p_T^\gamma$ bin. We plot $p p \rightarrow \gamma + \tchi^0 \tchi^0 $ so that there is no FSR in the Monte Carlo generated events as well, to give us a clean comparison of the analytical and MC generated events. In addition, we also plot the $p p \rightarrow \gamma + \tchi \tchi $ which includes both the neutralinos and charginos so that there is a photon in the ISR and FSR. We find that the a slightly larger fraction of the FSR photons have lower $p_T^\gamma$ than the ISR photons. In all cases, a cut on the $p_T^\gamma$ of the photon $>$ 10 GeV was imposed in order to cut off the collinear divergence. The effective operator estimation matches the falling photon $p_T^\gamma$ spectrum extremely well, to within a few percent.

One note about the viability of the effective operator paradigm in this case. One may raise the concern that the effective cut-off for the theory is low.  However, the difference in cross section between the exchange of a massive s-channel particle and a massless one is a simple pre-factor. In the case of the massive state it is $1/\Lambda^2$ and in the massless case it is $1/\hat{s}$, where $\hat{s}$ is the center-of-mass energy of the parton sub-process. We find that keeping $\hat{s}$ in the cross-section picks out higher $\hat{s}$ for larger photon $p_T^\gamma$ and therefore the fraction of events drops at higher $p_T^\gamma$. However, we find that this difference is negligible and we present our results with the simpler effective operator. In effect we are capturing only the kinematic information of the $2 \rightarrow 3$ process rather than the details about the intermediate state and overall production rate.

We can make the following general statement. For events with ISR bosons using the effective operator above, we expect the collinear divergence of the cross section to ensure that the production cross section is dominated by mono-bosons with very low $p_T$.  Such low $p_T$ events will be lost below any but the smallest standard model backgrounds, which are also dominated at low $p_T$.  Below, we will reinterpret existing ATLAS and CMS searches for this scenario and demonstrate the effectiveness of our analytical argument. Of course, the bottom line here is that mono-photon and mono-jet analyses will cut out most of the signal along with the background.

\subsection{Mono-jet and Mono-photon Search Sensitivities}
The ATLAS and CMS collaborations have used the mono-jet~\cite{ATLAS-CONF-2012-147, ATLAS:2012ky, Chatrchyan:2012me, CMS-PAS-EXO-12-048, CMS-PAS-SUS-13-009} and mono-photon~\cite{Aad:2012fw, CMS-PAS-EXO-12-047} final states to search for dark matter and large extra dimensions. These searches can be reinterpreted to obtain limits on the ewkino production cross-sections in the degenerate ewkino scenario. Our goal in this section is to scrutinize the results from our simple effective operator analysis by performing a full detector simulation. To this extent, we have implemented the selection criteria from two analyses: (i) 8 TeV ATLAS search for mono-jets in 10.5 fb$^{-1}$ of data~\cite{ATLAS-CONF-2012-147} (ii) 8 TeV CMS mono-photon search with 19.6 fb$^{-1}$ of data~\cite{CMS-PAS-EXO-12-047}. Both searches focus on events with a high $p_T$ leading jet or photon and hence are good candidates to test the validity of our analytical argument. The mono-jet search (left) and the mono-photon search (right) employ the following triggers and cuts:\\[5pt]

\begin{table}[h!]
\begin{center}
\renewcommand{\arraystretch}{1.0}
\scalebox{0.82}{
\begin{tabular}{|l|l|}
\hline
Search for mono-jets  & Search for mono-photons \\
\hline
$\bullet$ $\met  > $ 120 GeV & $\bullet$ $\met  > $ 140 GeV \\
$\bullet$ Leading jet with $p_T > $ 120 GeV and $| \eta| < $ 2.0 & $\bullet$ Photon with $p_T > 145$ GeV and $|\eta| < 1.4$ \\
$\bullet$ Veto events with $\ge 3 $ jets with $p_T > 30$ GeV and $|\eta| < $ 4.5 & $\bullet$ Veto events with $\ge$ one jet with $p_T > 30$ GeV \\
$\bullet$ The jet is required to be well-separated. & $\bullet$ The photon is required to be well-separated.\\
$\Delta \phi ({\rm jet}, \met) > 0.5$ & $\Delta \phi(\gamma, \slashed{E}_T) > 2$, $\Delta R(\gamma, \text{jet}) > 0.4$, $\Delta \phi(\text{jet}, \slashed{E}_T) > 0.4$.\\
$\bullet$ Events with an isolated lepton are rejected& $\bullet$ Events with isolated leptons are vetoed.\\
$p_T (e, \mu) > $ 10 GeV or $p_T (\tau) > $ 20 GeV & \\
\hline
\end{tabular}
}
\caption{\footnotesize{The selection criteria from existing mono-jet (ATLAS~\cite{ATLAS-CONF-2012-147}) and mono-photon (CMS~\cite{CMS-PAS-EXO-12-047}) searches. Parameters of the jet algorithm, photon and lepton isolation criteria can be found in the respective references.}}
\label{tab:cuts}
\end{center}
\end{table}

Both analyses binned the events into signal regions with varying $\met$ and the number of events in each signal region was found to be consistent with the background prediction from the SM. Neither search found any significant excesses in the 8 TeV data set and reported 95\% upper limits on events originating from new physics. We recast these searches in an attempt to constrain ewkino pair production by using the reported upper limits. We generate the $\tchi \tchi + j$ and $\tchi \tchi + \gamma$ events using \code{MadGraph5}~\cite{Alwall:2011uj}, shower them with \code{Pythia8.175}~\cite{Sjostrand:2007gs}, and use \code{PGS}~\cite{PGS} to perform the detector simulation.  The events are then filtered based on the selection criteria from the analyses (see~\ref{tab:cuts}) and compared to the reported observed numbers in each signal region.

\begin{table}[h!]
\begin{center}
\renewcommand{\arraystretch}{1.0}
\scalebox{0.66}{
\begin{tabular}{|c|c|c|c|c|c|c|c|}
\hline
 & $Z(\nu \nu) + j$ & $W (\tau \nu) + j$ & $W (\mu \nu) + j$ & $W (e \nu) + j$ & $\tilde{w}1 + j$ & $\tilde{h}2+j$ & $\widetilde{wh}3 +j$ \\
\hline
Production Cross section (in fb) &  $1.2 \times 10^6$ & $1.530 \times 10^6$ & $1.530 \times 10^6$ & $1.530 \times 10^6$ & $6.9 \times 10^3$ & $7.6 \times 10^2$ & $8.9 \times 10^2$ \\
\hline
Number of events at 10.5 fb$^{-1}$ & $1.26 \times 10^7$ & $1.6 \times 10^7$ & $1.6 \times 10^7$ & $1.6 \times 10^7$ & $7.24 \times 10^4$ & $8.00 \times 10^3$& $9.40\times 10^3$\\
$\slashed{E}_T > 120$ GeV & 327465& 197584 & 119202 & 118881 & 7368 & 1494 & 1605  \\
Jet $p_T > $ 120 GeV, $| \eta| < $ 2.0 & 196276 & 131302 & 71328 & 74541 & 4223 & 913& 970 \\
Veto events with 3 jets with & & & & & & &\\
$p_T > 30$ GeV \&  $|\eta| < $ 4.5 & 171964 & 99107 &62974 & 63296 & 3665&586 &609 \\
$\Delta \phi (\slashed{E}_T, j_1) > 0.5$ & 171964 & 83957 & 62974 & 63296 & 3665 &585 &609  \\
 Lepton Veto & 171964 & 83957 & 15743& 22491& 3665 & 548 & 539\\
 \hline
SR1 $\slashed{E}_T > 120$ GeV, Jet $p_T > $ 120 GeV & 173600 (171964) & 87400 (83957) & 34200 (15743) & 36700 (22491) & 3665& 548 & 539 \\
SR2 $\slashed{E}_T > 220$ GeV, Jet $p_T > $ 220 GeV & 15600 (17728) & 5580 (4103) & 2050 (1927) & 1880 (642) & 1057 & 173 & 176 \\
SR3 $\slashed{E}_T > 350$ GeV, Jet $p_T > $ 350 GeV & 1520 (1772) & 370 (315) & 158 (0) & 112 (0) & 210 & 40 & 33 \\
SR4 $\slashed{E}_T > 500$ GeV, Jet $p_T > $ 500 GeV & 270 (0) & 39 (0) & 42 (0)  & 16 (0) & 36 & 9.6 & 4.7 \\
\hline
\end{tabular}
}
\caption{\footnotesize{The table shows the cut-flow for the mono-jet analysis from the ATLAS collaboration~\cite{ATLAS-CONF-2012-147}. The background events are generated as a validation to the numbers reported by the collaboration. The number of background events reported by ATLAS can be found in the corresponding signal region (SR) rows and the numbers obtained from our simulation is quoted inside the brackets. To estimate the search efficiency, we use the numbers reported by the collaboration.}}
\label{tab:monojetcutflow}
\end{center}
\end{table}

SM backgrounds for the mono-jet process include W/Z + jets, $t\overline{t}$, single t, and QCD multijet production. We generate these backgrounds to validate our simulation tools and to estimate the reach of the analyses at the 13 TeV LHC. The cut flow for the background and the signal events for the mono-jet search are summarized in~\ref{tab:monojetcutflow} (for the three benchmarks - Wino 1, Higgsino 2 and Mixed 3 in~\ref{tab:Benchmarks}). The number of background events reported by ATLAS is also summarized in~\ref{tab:monojetcutflow} for each signal region (SR). These numbers can be compared with the background events obtained from our simulation (in parentheses) and we find good agreement between them. We note that our background estimates are limited by statistics and therefore to determine the search efficiency, we use the numbers reported by the collaboration. ATLAS observed the following number of events in each signal region: 350932 (SR1), 25515 (SR2), 2353 (SR3), 268 (SR4). Using the background prediction and the observed number of events, we calculate the 95\% Bayesian upper limit using a flat prior which, in the different signal regions translates to: 31508 (SR1), 1819 (SR2), 544 (SR3), and 39 (SR4).

From~\ref{tab:monojetcutflow} we that the wino-like point does the best with a sensitivity $S/\sqrt{B} \simeq 2$ (SR4) and just barely fails the reach of the ATLAS search, while the Higgsino and mixed states fail to reach sensitivity because of their cross-section which is much lower than the wino point. Our findings are in general, in agreement with the statements in~\cite{Baer:2014cua}, where in for the Higgsino LSP scenario $S/\sqrt{B}$ was generally only a few percent. The monojet search channel, however, is a promising search channel for wino-like LSPs at 13 TeV with $100 \ fb^{-1}$ where we find the $S/\sqrt{B} \gtrsim 5$ (SR4) for the Wino 1 benchmark point. Note, this channel may barely reach $S/\sqrt{B} \simeq 2$  sensitivity for predominantly Higgsino-like ewkinos (Higgsino 2 benchmark point) at the 13 TeV LHC. However, our background estimate here has large uncertainties. The case of Higgsino-like ewkinos at the 14 TeV LHC was however analyzed in Refs.~\cite{Han:2013usa, Baer:2014cua} with conflicting results. Ref.~\cite{Han:2013usa} claims a 2$\sigma$ significance with $\mu$ in the 100 - 150 GeV range and $3000 fb^{-1}$ integrated luminosity,  while Ref.~\cite{Baer:2014cua} argues that, since the signal and background have similar shapes and the events are significantly background dominated, it will be difficult to extract the signal even at high luminosity.

We perform a similar analysis to compare the mono-photon analysis from the CMS collaboration. Standard Model backgrounds for the mono-photon analysis are dominated by $Z\gamma$ production where the Z decays invisibly, and $W\gamma$ production with leptonic decay of the W. In~\ref{tab:cutphoton}, we show the cut-flow for the mono-photon search. For the mono-photon analysis, calculating the 95\% Bayesian upper limits using the 613 observed events, one finds that up to 123 events are allowed from any new physics that contributes to this channel. It is clear that our benchmark scenarios do not achieve sensitivity at 8 TeV. In general the best projection is for the wino-like scenario in which the sensitivity threshold $S/\sqrt{B}$ by a factor of 2.  Projections for mono-photon prospects at 13 TeV  with $100 \ fb^{-1}$ also fail to reach detectable sensitivity. Again our analysis is in line with estimates made by the authors of Ref.~\cite{Baer:2014cua} which were for Higgsino-like LSP scenario. Moreover, they conclude again that the signal to background ratio is too small for the signal to be extracted even at high luminosity.

\begin{table}[h!]
\begin{center}
\renewcommand{\arraystretch}{1.0}
\scalebox{0.86}{
\begin{tabular}{|c|c|c|c|c|c|}
\hline
 & $Z(\nu \nu) + \gamma$ & $W(l \nu) + \gamma$ &  $\tilde{w}1 + \gamma$ & $\tilde{h}2+\gamma$ & $\widetilde{wh}3 +\gamma$\\
\hline
Production Cross section (in fb) & $7.28 \times 10^3$ & $13.96\times 10^3$ &  $85.2$ & $12$ & $16$\\
\hline
Number of events at 19.6 fb$^{-1}$ & 142688& 273616 &  1669.9 & 235& 313\\
Photon $p_T > $ 145 GeV, $| \eta| < $ 1.4 & 473.7& 497.98 & 28 & 6.6 & 9.7\\
Missing Energy $>$ 140 GeV & 370.98 & 103.97 &  23 & 5.2  & 6.5 \\
$\Delta \phi (\gamma, \slashed{E_T}) > 2 $ & 370.98 & 98.5 & 23 & 5 & 6.4 \\
Jet veto if $p_T > 30$ GeV & 359.57& 82.08 & 21 & 4.1 & 5.3  \\
Lepton Veto & 359.57 & 49.2 & 21 & 3.7 & 4.5 \\
Events in signal region & 344.8 (359.57) & 102.5 (49.2) & 21 & 3.7 & 4.5\\
\hline
\end{tabular}
}
\end{center}
\caption{\footnotesize{The table shows the cut-flow for the mono-photon from the CMS collaboration~\cite{CMS-PAS-EXO-12-047}. The background events are generated as a validation to the numbers reported by the collaboration, which in some cases we are unable to match. The number of background events reported by CMS and the corresponding numbers obtained from our simulation is quoted inside the brackets.} }
\label{tab:cutphoton}
\end{table}

From our analysis and previous analyses of others it becomes clear that using the ISR or FSR signal of mono-jets or mono-photons for the discovery of ewkinos is quite difficult, due to the rapid fall off in $p_T$ of the signal for these events and the large backgrounds. This is also clear from the cut flow in~\ref{tab:cutphoton} where the sharp drop-off in the number of events is apparent with the high $p_T$ cut for the photon. With this in mind, we argue that a mono-Z signal with the Z decay into charged leptons provides a better option for probing ewkinos. Although the production cross-sections in this case are smaller,  the $p_T$ of the leptons are larger and thus, in some cases, more easily seen above the background.

\section {Mono-Z}
\label{sec:monoz}
As we saw from the previous section we cannot expect that any process dominated by the radiation of a collinear jet or photon will be a useful discovery channel for new stable uncharged particles, ewkinos or otherwise, due to the shape of the $p_T$ distribution. However if the particle needed to trigger on, which was radiated in the initial state was massive, we might expect to find its decay products. We thus propose mono Z + $\met$ as a suitable discovery channel for ewkino pair production. In particular we propose the leptonic final state channel of Z decay as the most promising channel. The relevant event topology is thus, $pp \rightarrow Z \tchi\tchi \rightarrow \ell\ell \tchi\tchi$.

The leptonically decaying Z offers two advantages over mono-jet and mono-photon analyses.  One is that the background for the 2 lepton $+$ $\met$ final state is very low. The second is that though we expect the Z to be produced with low $p_T$, in its decay it will impart to the leptons substantial momentum and provide a possible trigger. ATLAS has used the mono-Z final state topology using the 8 TeV, 20.6 fb$^{-1}$ dataset~\cite{Aad:2014vka} to look for dark matter models in final state with one Z boson decaying to leptons plus massless particles which appear as missing energy. This search can easily be reinterpreted as a search for new ewkino sector physics with the mono-Z final state. The ATLAS search has the following event selection criteria:

\begin{itemize}
\item Two same-flavor opposite-sign electrons or muons, each with $p_{\rm T}^{\ell} > 20$ GeV, $|\eta^{\ell}|<2.5$;
\item Di-lepton invariant mass close to the $Z$ boson mass: $m_{\ell\ell} \in [76, 106]$ GeV;
\item No particle level jet with $p^j_{\rm T} >$ 25 GeV and  $|\eta^j|<$4.5;
\item $p_{\rm T}^Z /\met > 0.5$;
\item $ \Delta \phi (p_{\rm T}^Z, \met) > 2.5$
\end{itemize}

Finally, the events that pass all the above cuts are binned into 4 different distinct signal regions:
\begin{itemize}
\item $\met > 150,\ 250,\ 350,\ 450 $ GeV  constituting 4 signal region bins
\end{itemize}

The Standard Model background to a mono-Z search is mainly the SM production of ZZ where one Z decays invisibly and the other leptonically, or to SM WZ production with lost or misidentified lepton. Once again, we use \code{MadGraph5} to generate events at 8 and 13 TeV, shower them using \code{Pythia} and perform detector simulation using \code{PGS}. We compare the 8 TeV SM backgrounds with the numbers presented in the ATLAS analysis to validate our tools and in general find good agreement. The first step is to reinterpret the 8 TeV analysis.

\begin{table}[h!]
\begin{center}
\renewcommand{\arraystretch}{1.0}
\scalebox{0.86}{
\begin{tabular}{|c|c|c|c|c|c|}
\hline
 & $Z(\nu \nu) Z(l^+ l^-) $ & $W(l \nu) Z(l^+ l^-) $ & $\tilde{w}1 + Z$ & $\tilde{h}2+Z$ & $\widetilde{wh}3 +Z$\\
\hline
Production Cross section (in fb) & 93 & 98 & 24 & 39 & 158 \\
\hline
Number of events at 20.3 fb$^{-1}$ & 1893 & 1989 & 487 & 791 & 3207\\
Events with a Z ( $m_{ll} \epsilon [76, 106]$ ) & 1286 & 1358 & 18 & 30.9 & 130\\
$\Delta \phi (p_T^Z, \slashed{E_T}) > 2.5 $ & 1112 & 810 & 15 & 22& 85 \\
$p_T^Z/\slashed{E_T} > 0.5$ & 955 & 507 & 13 & 17 & 62 \\
Jet veto if $p_T > 25$ GeV & 870& 421 & 11 & 8 & 36 \\
Lepton Veto & 869 & 421 & 11 & 7.9 & 33 \\
\hline
SR1 ($\slashed{E_T} > 150$ GeV) & 41 (38) & 8.0 (6.4) & 1.75 & 2.05 & 4.1 \\
SR1 ($\slashed{E_T} > 250$ GeV) & 6.4 (6.0) & 0.8 (0.8) & 0.43 & 0.55 & 0.64 \\
SR1 ($\slashed{E_T} > 350$ GeV) & 1.3 (1.25) & 0.2 (0.51) & 0.04 & 0.07 & 0 \\
SR1 ($\slashed{E_T} > 450$ GeV) & 0.3 (0.22) & 0.1 (0.39) & 0 & 0 & 0 \\
\hline
\end{tabular}
}
\caption{\footnotesize{Cutflow for the mono-Z analysis from ATLAS~\cite{Aad:2014vka}.The background events are generated to validate our simulation tools (quoted inside the brackets) and they match the expected SM backgrounds reported by the collaboration.}}
\label{tab:monoz8}
\end{center}
\end{table}

In~\ref{tab:monoz8} and~\ref{monoZ13} we show cut flow chart for three ewkino benchmark scenarios along with the SM backgrounds. \ref{tab:monoz8} shows the reinterpretation of the ATLAS 8 TeV analysis~\cite{Aad:2014vka}, while in~\ref{monoZ13} we implement the same cuts to to study the reach of the mono-Z channel at 13 TeV with $100 \ fb^{-1}$ of data.

\begin{table}[h!]
\begin{center}
\renewcommand{\arraystretch}{1.0}
\scalebox{0.86}{
\begin{tabular}{|c|c|c|c|c|c|}
\hline
 & $Z(\nu \nu) Z(l^+ l^-) $ & $W(l \nu) Z(l^+ l^-) $ & $\tilde{w}1 + Z$ & $\tilde{h}2+Z$ & $\widetilde{wh}3 +Z$\\
\hline
Production Cross section (in fb) &  162 & 160.9 & 56 & 102.3 & 376\\
\hline
Number of events at 100 fb$^{-1}$ & 16200 & 16200 & 5600 & 10230 & 37600\\
Events with a Z ( $m_{ll}\ \epsilon\ [76, 106]$ ) & 10949 & 11014 & 199 & 394 & 1248\\
$\Delta \phi (p_T^Z, \slashed{E_T}) > 2.5 $ & 9365 & 6555 & 175 & 290 & 887 \\
$p_T^Z/\slashed{E_T} > 0.5$ & 7926 & 4110 & 155 & 208 & 571 \\
Jet veto if $p_T > 25$ GeV & 7073& 3357 & 118 & 108 & 353 \\
Lepton Veto & 7072 & 1948 & 118 & 97 & 319 \\
\hline
SR1 ($\slashed{E_T} > 150$ GeV) & 369 & 66.7 & 28.5 & 25 & 22.56 \\
SR1 ($\slashed{E_T} > 250$ GeV) & 61 & 7.7 & 11.7 & 4.09 & 0 \\
SR1 ($\slashed{E_T} > 350$ GeV) & 15 & 1.62 & 2.2 & 1.02 & 0 \\
SR1 ($\slashed{E_T} > 450$ GeV) & 2.5 & 0.97 & 1.1 & 1.02 & 0 \\
\hline
\end{tabular}
}
\end{center}
\caption{Projections for the mono-Z search at 13 TeV LHC with $100 \ fb^{-1}$ based on the cuts from ATLAS~\cite{Aad:2012fw}.}
\label{monoZ13}
\end{table}

\begin{figure}[h!]
\begin{center}
\subfigure[\footnotesize Wino LSP]{
\includegraphics[width=7.67cm]{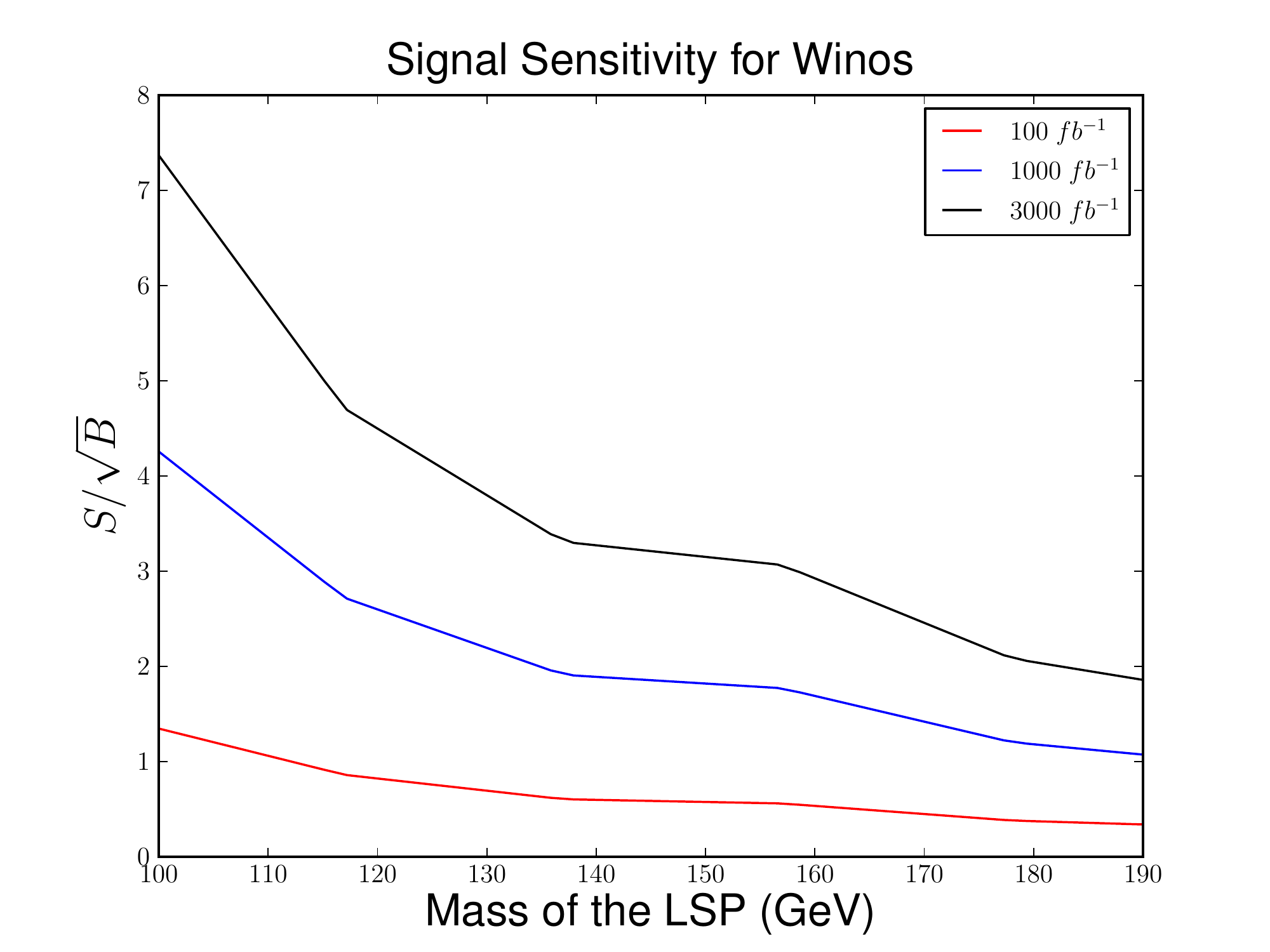}
}
\subfigure[\footnotesize Higgsino LSP ]{
\includegraphics[width=7.67cm]{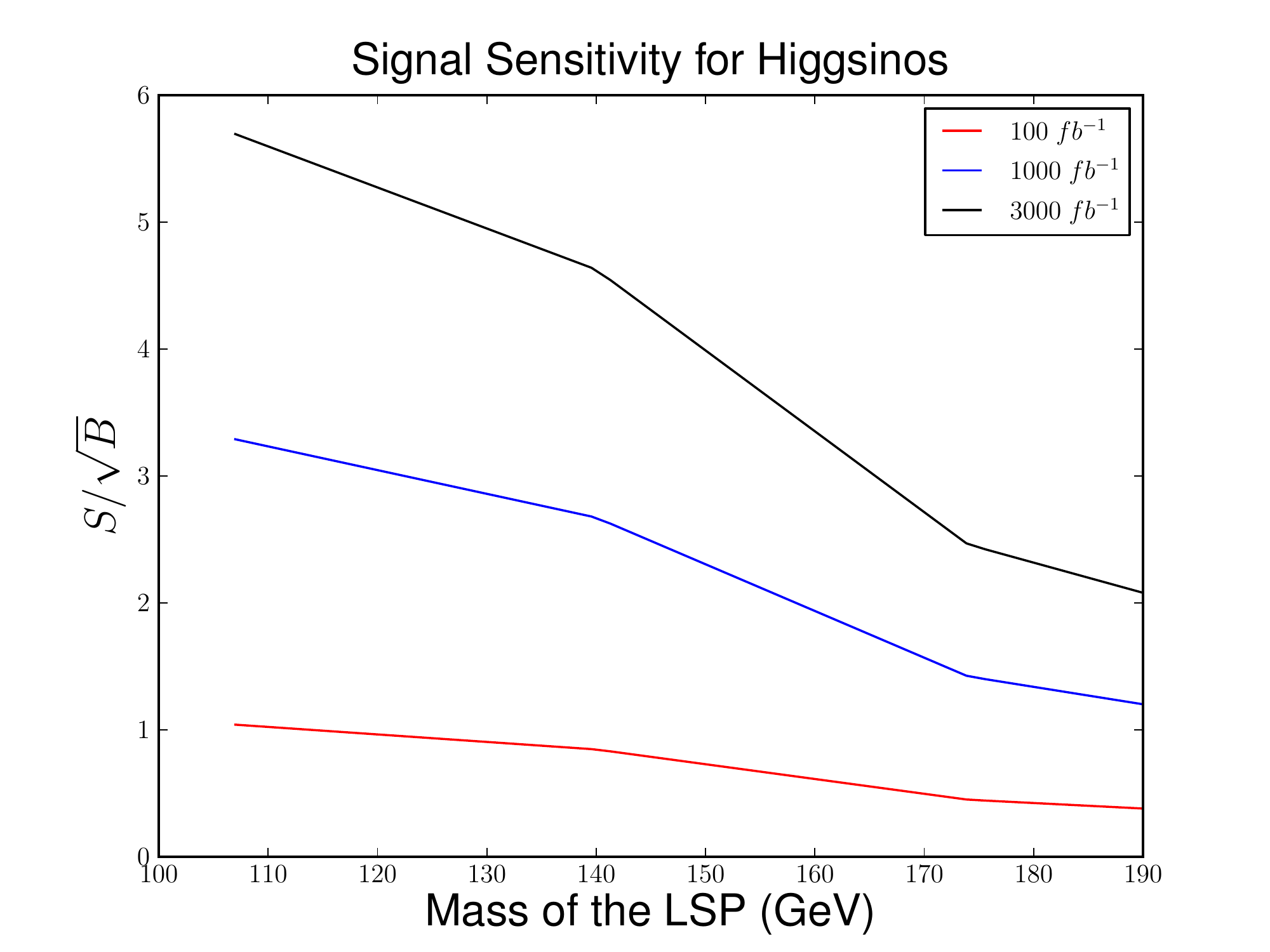}
}
\newline
\subfigure[\footnotesize Mixed wino-higgsino LSP]{
\includegraphics[width=10cm]{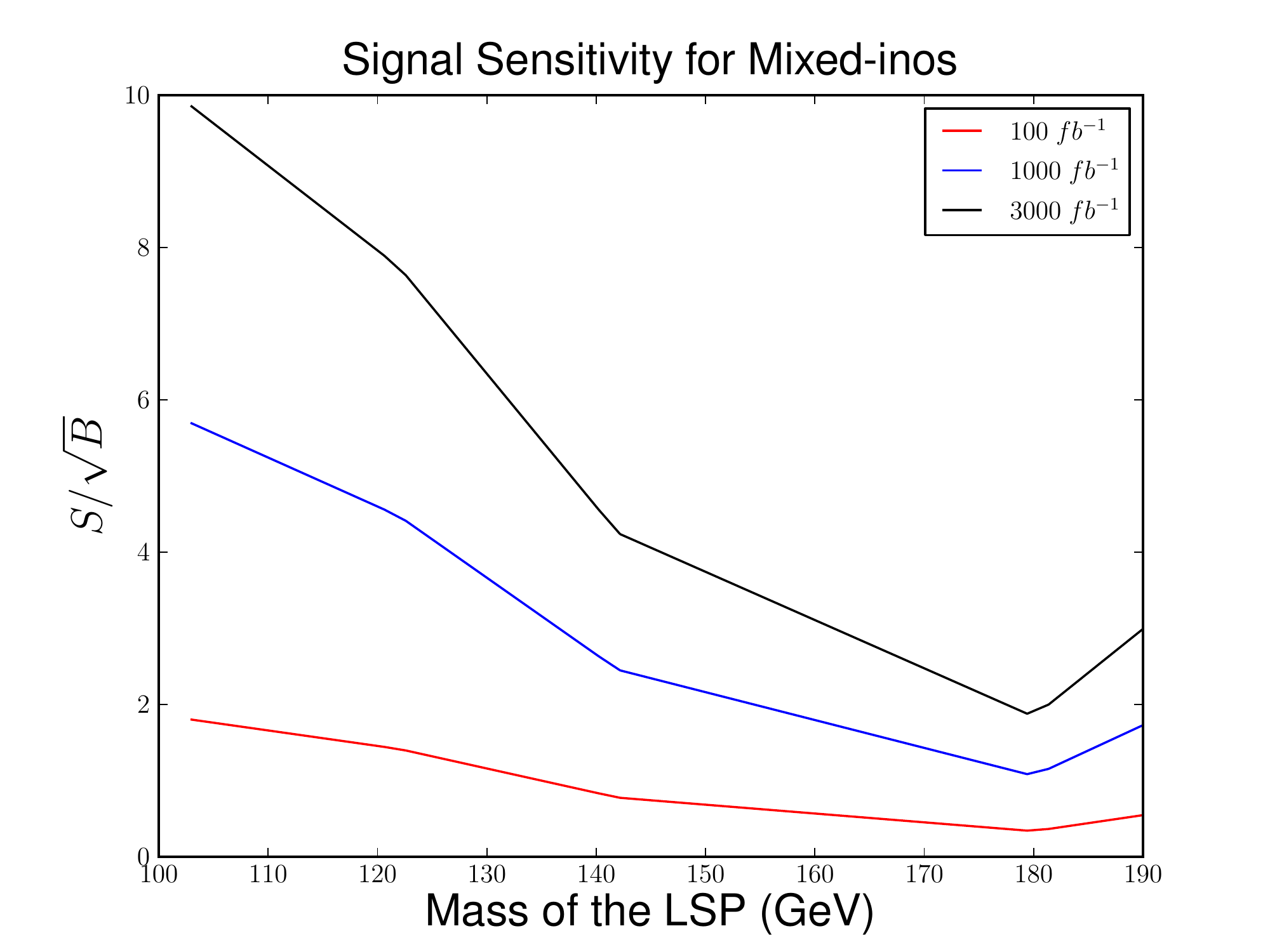}
}
\caption{Signal sensitivity of the mono-Z search for wino, higgsino and mixed ewkino pair productions.}
\label{fig:sovb}
\end{center}
\end{figure}

We show that the mono-Z searches have a very good signal to background ratio. Although there is no sensitivity with the 8 TeV analysis, the degenerate ewkino scenario fares well in the 13 TeV projection.  This can be seen in~\ref{fig:sovb} where we show the LHC sensitivity reach to wino-like, higgsino-like, and mixed ewkino scenarios in the mono-Z channel.  We present results at $100\ fb^{-1}$,$1\ ab^{-1}$ and $3\ ab^{-1}$, for a range of LSP masses.  We see that with $3\ ab^{-1}$ of data LHC has 5-sigma discovery potential to mixed ewkinos under 140 GeV, and even to light pure winos and higgsinos up to 120 GeV, and maintains 2-3 sigma sensitivity in all degenerate ewkino scenarios up to 200 GeV.  In the mixed case, $1\ ab^{-1}$ of data affords sensitivity across the entire 100-200 GeV mass range.

Note, however, for the wino benchmarks, mono-jets and mono-Z are about the same. One is not better than the other. In fact when the LSPs are wino-like, the mass difference is always very small (see~\ref{deltam}) since $\mu$ is large for wino.). Hence for wino-like scenarios, other strategies like disappearing charged tracks may be the best search option. However, where mono-jets and mono-photons fail for the higgsino and the mixed scenarios, we expect the mono-Z search to have a strong sensitivity. For these cases, $\Delta M > 1$ GeV and all the other search options (such as track differentiation) will likely fail.

\section{More complex topologies}
\label{sec:other}
\begin{figure}[h!]
\begin{center}
\includegraphics[width=10cm]{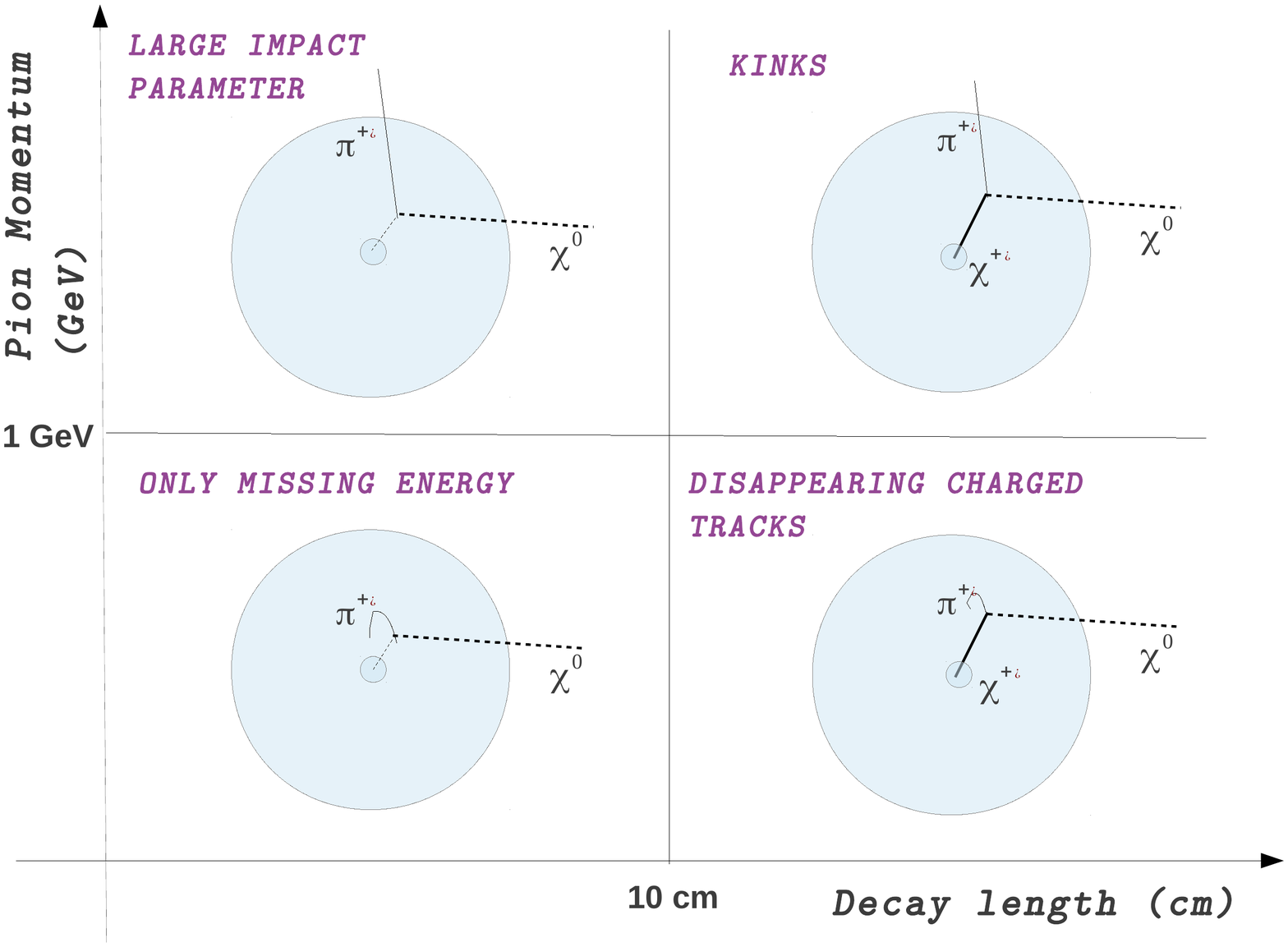}
\caption{\label{topologies} Possible chargino decay topologies in different regimes with varying chargino lifetimes. The long lived charginos predominantly decay into a single pion $+$ neutralino.  For promt chargino decays, other decay modes open as well, nevertheless, the decay products remain soft and appear as only missing energy to the detector.}
\end{center}
\end{figure}

For very small mass splittings in the degenerate ewkino scenario, those under 1 GeV, the charginos may be quite long lived. In this regime the event topology becomes quite complex , see for example~\cite{Feng:1999fu}. In~\ref{tab:Benchmarks}, we compiled a series of benchmark points for Wino-like, Higgsino-like and well-mixed chargino scenarios to demonstrate mass splittings and chargino lifetimes. Notice that in the case of pure wino, the mass splittings are generally much smaller since $\mu$ is typically large. In the case of pure Higgsinos, the mass splitting is then inversely proportional to the difference between $M_1$ and $M_2$. On the other hand, for mixed-state ewkinos, the mass difference is typically greater than 1 GeV. Below we enumerate the possible event topologies in order of decreasing mass splitting.   In~\ref{topologies}, we have constructed a diagrammatic picture of these possible decays.  The diagram shows the different final state topologies as a function of the chargino decay length and momentum of final state decay products.

For mass splittings between 1 and 10s of GeV, the decay of the chargino is prompt and the decay signature is missing energy $+$ soft jets, $\gamma$ or Z as discussed in this work. For $\Delta M < 1$ GeV the decay is not prompt, and the chargino will travel micro-meters or more. At these small mass splitting, the chargino decay will proceed predominantly through the single pion mode, $\tchi^\pm \rightarrow \tchi^0 + \pi^\pm$. If the chargino lifetime is macroscopic, yet not large enough to penetrate sufficiently into the detector's central tracker before it decays, the decay will appear either as a pion track with a large impact parameter, or pure missing energy depending on the momentum of the decay products. If the chargino penetrates several layers of the detector's tracker it may leave a detectable track. Then if its decay product, the pion has sufficient momentum the event will appear either as a kink, otherwise it will be a disappearing track. ATLAS has performed a disappearing charged track search~\cite{Aad:2013yna} with 20 fb$^{-1}$ of data at 8 TeV and have been able to be significant constraints on the degenerate region where the chargino lives long enough to leave cm long tracks. Using this search, ATLAS was able to rule out up to $ \simeq 500$ GeV charginos with lifetime of 10 ns and up to 170 GeV charginos with lifetime of 0.1 ns. Notice that this already rules out the wino 1 and the Higgsino 1 benchmark point considered in the paper. This demonstrates that the disappearing charged track searches will provide a background free, clean strategy for the very degenerate scenarios.

The large impact parameter and the kinks have not been implemented by the collaborations yet. Triggering on a isolated pion may present a challenge for implementing these searches. The mono-Z search proposed in this work can be applied to all sufficiently small mass splittings, and thus covers the parameter space of all 4 search topologies above. The mono-Z handle may still have something to offer these non-standard search strategies.  In the topologies  where charginos may decay with a displaced vertex, initial state radiation is still necessary as a search trigger. Current searches~\cite{Aad:2013yna} rely on mono-jet triggers, but we note that following our earlier arguments about event $p_T$, we expect the inclusion of mono-Z to be a clean low background ISR trigger for topologies with displaced vertices.

\section{Conclusions}
\label{sec:conclusion}
We find that in the SUSY scenario of mass degenerate charginos and neutralinos, the mono-Z search channel is a viable pathway for detection and discovery at the high luminosity LHC. We have presented general arguments why mono-Z searches may succeed where mono-jet and photon-searches will always give extremely soft radiation on which to trigger.

The degenerate ewkino scenario in SUSY is determined by the values of $M_1$, $M_2$ and $\mu$ and these parameters set the mass difference and the ewkino content. Wino LSPs predominantly prefer $\Delta M \leq 500 MeV$ and therefore will be tested by sophisticated techniques such as disappearing charged tracks. On the other hand, the higgsino and especially the mixed ewkino scenarios could have $\Delta M$ in the range of few 100 MeV to 10s of GeV and the signal will be indifferentiable from missing energy. In these cases, the mono-boson ISR/FSR trigger along with large $\met$ provides a method of constraining these models. It has been shown over and over again now that backgrounds dominate the mono-jet searches for ewkinos and we showed in our calculations that the mono-photons will not pass the high photon $p_T$ cuts due to the collinear divergence of the amplitude in the low photon momentum. We strongly advocate the use of mono-Z triggers where the on-shell Z would impart enough momentum to the leptons to pass the lepton $p_T$ triggers. In addition, the mono-Z channel is relatively less contaminated with background events. For Higgsinos and mixed ewkinos, this channel is quite promising.

We note these results are applicable not only to the ewkino scenario but also to other searches with mono-boson plus $\met$ triggers. The mono-Z channel may be effective for ruling in or out more general degenerate SUSY scenarios.  For example this method could be quite effective for ruling out scenarios with degenerate squark and neutralino masses, these scenarios have the benefit of over-all larger pair production cross sections of SUSY states. Further, we expect that following our arguments, mono-W events, may also be a possible viable search signature for ewkino pair production.  We have also demonstrated the effectiveness of a simple effective operator technique for capturing the most important kinematic features of SUSY pair production scenarios with an ISR photon.

\section*{Acknowledgements}
The authors would like to thank Christopher Hill and Charles Bryant for useful discussions. We thank the Ohio Supercomputer Center for their computing resources. AA is funded through the Ohio State University Presidential Fellowship.  AA and SR acknowledge partial support from DOE grant DOE/ DE-SC0011726.   Also SR thanks the Mainz Institute for Theoretical Physics, part of the Excellence Cluster "PRISMA," located on the campus of the Johannes Gutenberg-University, Mainz, Germany for partial support during the final preparation of the paper.

\clearpage

\newpage

\clearpage
\newpage

\bibliography{bibliography}

\bibliographystyle{utphys}

\end{document}